# On the *E2* admixture in the deexcitation of the $^{229m}$Th isomer

F. F. Karpeshin[1], M. B. Trzhaskovskaya[2] and L. F. Vitushkin[1]

[1]*D. I. Mendeleyev Institute for Metrology*

[2]*National Research Center "Kurchatov Institute" ─ Petersburg Nuclear Physics Institute*

Latest experimental data on the properties of $^{229}$Th isomer nuclide are analyzed. Deduced are the values of the lifetime and the *E*2-fraction in its electromagnetic deexcitation, based on the latest experimental data concerning the isomer: its energy together with the magnetic and quadrupole moments.

## 1. Introduction

Study of the properties of the isomeric level in $^{229}$Th is a topical problem. Interest is caused by its uniquely low energy, which is within ten eV [1-6]. A summary of its experimental values is given in Section 7. For definiteness, we will use a value of 7.6 eV, as reported in Ref. [2], in the calculations below, unless otherwise specified. Within the uncertainties, it does not contradict the rest of the measurements [1,3-6]. $^{229}$Th nuclide is the most likely candidate for the creation of next-generation nuclear-optical clock [7]. It has the lowest excitation energy among the known nuclei. With such energy, the level is "entangled" with many atomic levels. For this reason, it undergoes deexcitation mainly through internal conversion (IC) in neutral atoms, and electron bridges in ions [2]. Below we analyze experimental data recently obtained in the following milestone works:

1) The discovery by the Munich group of the conversion decay from the isomeric state [3], predicted back in [8]. The lifetime of the isomer exactly coincided with the predicted one: 10 $\mu$s.

2) Measurement at the Physikalisch-Technische Bundesanstalt (PTB) in Braunschweig of the magnetic dipole and electric quadrupole moments of the $^{229}$Th nucleus in the isomeric state [9]. The following values in the ground state were previously known: the magnetic moment $\mu = 0.360$ (7) $\mu_N$ ($\mu_N$ is the nuclear magneton), and the internal quadrupole moment $Q_0 = 8.8\ (1)\ e$b. The moments in the isomeric state appeared to be $\mu_m = -0.37\ (6)\ \mu_N$, and $Q_0^m = 8.7\ (3)\ e$b. Let us analyze these two experimental results in more detail.

## 2. Discovery of the direct IC decay of the isomer

It was in 2016, that direct decay of the isomer to the ground state through IC was observed for the first time at the Ludwig and Maximilian University of Munich (LMU) [3]. In these measurements, isomeric recoil nuclei were produced in the alpha decay of $^{233}$U. The isomeric half-life in neutral atoms was measured, which turned out to be equal to 7 $\mu$s. The question arises: what information on the properties of the isomer can be retrieved from this result. Some answers were given in Ref. [10]. Let us consider this question in more detail here.

The life time taking into account IC is defined as

$$\Gamma_c(M1) = (1 + \alpha(M1))\Gamma_\gamma^{(n)}(M1), \qquad (1)$$

where $\alpha(M1) = 1.29 \times 10^9$ [8] is the internal conversion coefficient (ICC). The unity in (1) can be neglected. Depending on the energy of the isomer $\omega_n$, the ICC and the radiative nuclear width behave differently:

$$\Gamma_\gamma^{(n)}(M1) \sim \omega_n^3,$$

$$\alpha(M1) \sim \omega_n^{-3}.$$

Therefore, the lifetime turns out to be independent of the energy of the isomer. By comparing with experiment, one can determine the only characteristic: the Weisskopf's prohibition factor $H$ [11]. Making use of the reference material presented in paper [12], one arrives at the Weisskopf hindrance factor of $H = 80$, which corresponds to the above half-life of 7 $\mu$s. This can be considered as the value directly following experiment [3].

This experiment became a starting point for the next important experiment [9], carried out by the combined efforts of the PTB and LMU groups. In this experiment, the above values of the moments of the nucleus in the isomeric state were determined. The measurements were based on the dependence of the hyperfine structure (HFS) on the nuclear moments. In experiment [9], the HFS of two electronic levels was studied in doubly ionized Th$^{++}$ atoms: those of 63 and 20711 cm$^{-1}$. Comparative measurements were performed in neutral $^{229}$Th atoms. Using known values of the nuclear moments in the ground state, the moments in the isomeric state have been determined.

Since then the LMU group further developed the equipment, and this made it possible to separate the conversion electrons from the residual ions, and to measure the spectrum of the conversion electrons. As a result, the value of the isomer energy given below has been obtained [4]. It turned out to be within the error bars of the previous measurements [1,2]. Moreover, in the last minute, we got aware of the latest measurement of the isomer energy in Heidelberg [13], also in agreement with the Munich value [4].

## 3. Analysis of the nuclear moments in the $^{229m}$Th isomeric state

Use of the measured values of the magnetic and quadrupole moments $^{229}$Th in the isomeric state allows us to essentially advance our knowledge of the properties of the nuclide. First of all, knowledge of the quadrupole moment $Q_0$ allows one to calculate the transition intensities between the states belonging to the rotation band starting at the isomeric level $I^\pi K = 3/2^+3/2$. Specifically, this concerns the 29.19-keV transition from the first excited $5/2^+3/2$ to the isomer bandhead, see Figure.

There are four independent partial transitions from this state: intraband $M1$ and $E2$ transitions to the isomer state, and two interband transitions to the ground state. Furthermore, radiative transitions induce IC ones, much more intense. Let the *in* and *cr* superscripts indicate the intra- and interband transitions from the 29.19 keV level; the subscripts $\gamma$, $c$, and $t$ indicate that they refer to the radiative, conversion, or full width, respectively. Making use of the quadrupole moment in the isomer state [9] in combination with the total width of the level measured in recent work [5], one can refine the transition intensities from this state, and finally draw a conclusion about the lifetime of the isomeric level and the multipole mixture of its decay. Let us consider such analysis in more detail.

**Quadrupole intraband $E2$ transitions.** Here and below, the *in* and *cr* superscripts indicate the intra- and interband transitions; the subscripts $\gamma$, $c$, and $t$ indicate that they refer to the radiative, conversion, or full width, respectively.

The reduced probability of the radiative intraband *E2*-transitions is determined by means of the following formula [11]:

$$B(E2; I_1 K \to I_2 K) = \frac{5}{16\pi} Q_0^2 C^2(I_1 K 20 | I_2 K). \tag{2}$$

In (2), $C(I_1 K 20 | I_2 K)$ – the Clebsch–Gordan coefficients, $I_1$, $I_2$ are the spins of the nucleus in the initial and final states, and *K* is the projection of the internal angular momentum onto the axis of symmetry of the nucleus. In turn, the partial probability of the radiative transition of multipolarity $\tau L$ per unit time $\Gamma(\tau L)$ is related to the reduced probability $B(\tau L)$ as follows:

$$\Gamma(\tau L) = \frac{8\pi(L+1)}{L[(2L+1)!!]^2} \omega_n^{2L+1} B(\tau L). \tag{3}$$

Using formulas (2) and (3), we can calculate the probability of the interband transition:

$B(E2) = 1.28 \ e^2 b^2$, and accordingly $\Gamma_\gamma^{in}(E2) = 2.2 \times 10^{-10}$ eV.

ICC values for this transition calculated by the RAINE program [14] are $\alpha(M1) = 139.8$, $\alpha(E2) = 4270$. Thus we arrive at the total partial width of the transition

$$\Gamma_t^{in}(E2) = \Gamma_\gamma^{in}(E2) + \Gamma_c^{in}(E2) = (1+\alpha(E2))\Gamma_\gamma^{in}(E2) = 0.939 \times 10^{-6} \text{ eV}.$$

***M1* intraband transitions.** The magnetic moment $\mu$ of a collective rotational state with quantum numbers *IK* is determined by the formula

$$\mu = g_R I + (g_K - g_R)\frac{K^2}{I+1} \tag{4}$$

with the parameters $g_R$ - gyromagnetic ratio for the collective rotation of the nucleus as a whole, and $(g_K - g_R)K$, characterizing the internal motion of nucleons in the coordinate system associated with the nucleus. The probability of the radiative interband *M1* transitions is fully determined by the internal motion of the nucleons:

$$B(M1; I_1 K \to I_2 = I_1 \pm 1, K) = \frac{3}{4\pi} \mu_N^2 (g_K - g_R)^2 K^2 C^2(I_1 K 1 0 | I_2 K). \qquad (5)$$

If the first term could be neglected in (4), knowing the magnetic moment would be sufficient to calculate the probability of interband transitions. However, in reality, the contribution of the collective rotation to (4) dominates. Therefore, knowledge of the magnetic moment is not enough, in order to determine the probability of the intraband *M1* transitions, in contrast to the quadrupole ones. The parameter $(g_K - g_R)^2 K^2$ can be determined from the known probabilities of interband radiative transitions between higher-lying states of the nucleus, which are populated, for example, in the $^{233}$U $\alpha$ decay [15] or the $^{229}$Ac $\beta$ decay [16]. Calculations in paper [15] were performed in the strong coupling rotational model under assumption that the Coriolis interaction plays no significant role. In particular, in this way, a value of $B(M1) = 0.048 \mu_N^2$ for the 29.19-keV *M1* interband transition to the isomeric state was obtained. If we neglected the collective term in (4), we would have obtained a value 0.238 $\mu_N^2$ for *B (M1),* i.e., five times as much. This shows the dominant role of the collective rotation in formation of the magnetic moment of the isomeric state. Using formula (3), we obtain an estimation of the partial radiative width of the 29.190-keV transition to the isomeric state $\Gamma$ *(M1)* = $1.38 \times 10^{-8}$ eV.

Similarly, the reduced probabilities of the other interband transitions were calculated in [15], including the isomeric transition $5/2^+3/2$ to $5/2^+5/2$, whose energy was considered to be of 3.5 eV. However, questions remain in this way, related to the necessity of separating the contributions from the *M1* and *E2* transitions. The problem is complicated due to the influence of the Coriolis interaction, phonon interaction, non-adiabatic effects and other factors. These factors were taken into account when analyzing the multipole mixture of radiative transitions in a later paper [16]. A comparison of the results [15,16] shows significant differences, which indicate the importance of taking the above effects into account.

**Interband transitions.** For our purposes, study of the closest interband transition of 29.19 keV $5/2^+3/2$ to $5/2^+5/2$ is most essential. For the first time, direct information on the radiative width of this transition was provided by a recent experiment [5]. The photoexcitation cross section was measured to the level of 29.19 keV $5/2^+3/2$. The innovation of this method is evident: it is based on a direct measurement of the partial radiative width of the transition from the ground state $5/2^+5/2$ to the excited state, belonging to the isomeric rotational band. Moreover, the electron shell does not affect the probability of photoexcitation. Therefore, there is no gain in favor of the *E2* component, as in the case of radiation deexcitation in neutral atoms due to large values of ICC for this transition. Consequently, we can conclude that the role of the admixture of the *E2* component in the photoexcitation cross section is negligible. And one can attribute the width $\Gamma_\gamma^{cr}(M1) = 1.70 \pm 0.40$ neV measured in this work to the pure radiative width of the *M1* transition $5/2^+3/2 \rightarrow 5/2^+5/2$. Taking IC into account, we arrive at the full width of the *M1* component of this transition:

$$\Gamma_t^{cr}(M1) = (1+\alpha(M1))\Gamma_\gamma^{cr}(M1) = 0.238 \times 10^{-6} \text{ eV}.$$

Furthermore, the half-life of the level of 29.19 keV was also measured in [5]: $T_{1/2} = 82.2 \pm 4$ ps, corresponding to the full width $\Gamma_t = (5.55 \pm 0.04) \times 10^{-6}$ eV. In fact, these data are sufficient for a direct estimation of the characteristics of both the interband and intraband transitions from this level, as well as the isomeric transition in the framework of the collective rotation nuclear model. Let us see what restrictions this imposes on the reduced probability of the intraband *M1* transition. The full width consists of the four components:

$$\Gamma_t = \Gamma_t^{in}(M1) + \Gamma_t^{in}(E2) + \Gamma_t^{cr}(M1) + \Gamma_t^{cr}(E2). \qquad (6)$$

In the first approximation, the width of the interband *E2* transition $\Gamma_t^{cr}(E2)$ can be neglected as compared to $\Gamma_t^{in}(E2)$. Then from (6) we immediately obtain the value $\Gamma_t^{in}(M1) = 4.37 \times 10^{-6}$ eV, and, taking into account IC, the proper nuclear radiative width $\Gamma_\gamma^{in}(M1) = 3.12 \times 10^{-8}$ eV. Thus, the total width of the level of 29.19 keV is nearly exhausted by the intraband *M1* transition (see Figure). Admixture of the *E2* transitions is characterized with the parameter $\delta^2 = \Gamma_\gamma^{in}(E2) / \Gamma_\gamma^{in}(M1)$, whose more accurate value is listed below.

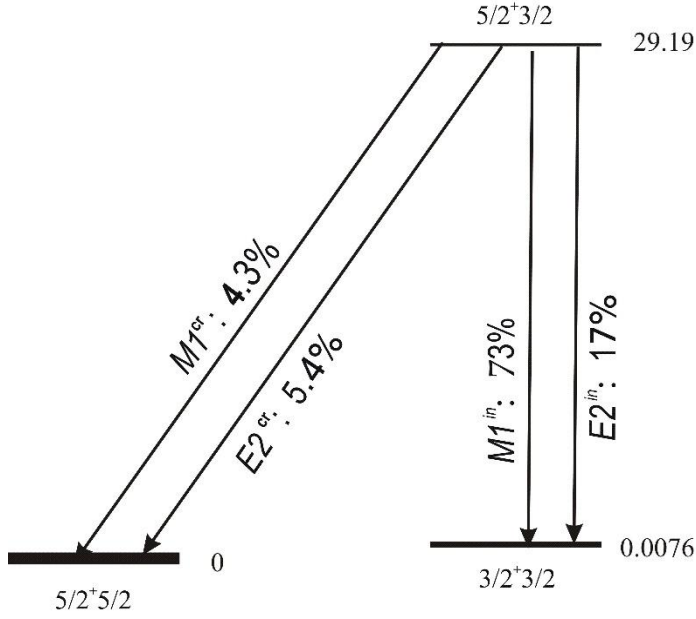

Fig. The *M1* and *E2* transition scheme between the ground and isomeric levels, on the one hand, and the 29.19 keV rotational level 5/2⁺3/2, built on the isomeric state as the bandhead, on the other. The full intensities are presented, allowing for the radiative and conversion transitions in neutral atoms of $^{229}$Th. They are normalized to 100%. Full width of the 29.19-keV state is 5.55×10⁻⁶ eV [5].

One can further amend the above values, explicitly taking into account the interband *E2* transition. Maximal experimental value for this component follows from [16]: $B(E2) = 0.41\ e^2\ b^2$, whence $\Gamma_\gamma^{cr}(E2) = 7.05\times10^{-11}$ eV, and и $\Gamma_t^{cr}(E2) = 0.301\times10^{-6}$ eV. As a result, the total width of the intraband *M1* transition in (6) only slightly decreases and becomes $4.07 \times 10^{-6}$ eV. Taking into account the ICC, we find the radiation width $\Gamma_\gamma^{in}(M1) = 2.91\times10^{-8}$ eV.

For clarity, we present the values obtained above in the figure. The admixtures of the *E2* components are 0.0076 and 0.041 for the intraband and interband transitions. Taking into account the conversion channel, the admixture parameters of the *E2* component become $(\delta_c^{in})^2 = \Gamma_c^{in}(E2)/\Gamma_c^{in}(M1) = 0.23$, and $(\delta_c^{cr})^2 = 1.26$.

## 4. Lifetime of the isomeric state of $^{229m}$Th and multipole mixture of its decay

Now we take the decisive step: exploiting the reduced widths for the 29-keV interband transition, we calculate the reduced probabilities *B(E2)* and *B (M1),* and the lifetime for the isomeric state by means of formulas (2), (5). In this section, we assume a value of 8.2 eV as the transition energy. Radiative width of the main *M1* component can be obtained from $\Gamma_\gamma^{cr}(M1)$, taking into account proportionality of the rate to the energy $\omega^3$, and multiplying the result by the ratio of the squares of Clebsch–Gordan coefficients $C^2$ (5/2 3/2 10 | 5/2 3/2) / $C^2$ (3/2 3/2 11 | 5/2 5/2) = 3.5. As a result, we arrive at $\Gamma_\gamma^{is}(M1) = 1.32 \times 10^{-19}$ eV. The corresponding half-life is one hour. Similarly, for the admixture of the *E2* component we obtain $\Gamma_\gamma^{is}(E2) = 1.23 \times 10^{-28}$ eV. This implies the estimation for the mixing parameter $\delta^2$ of the isomeric transition: $\delta^2 = 0.93 \times 10^{-9}$. The calculated total ICC for this transition is $\alpha\ (M1) = 1.11 \times 10^9$, $\alpha\ (E2) = 8.46 \times 10^{15}$. Taking into account the ICC, we find the multipole widths of conversion transitions $\Gamma_c^{is}(M1) = 1.47 \times 10^{-10}$ eV, $\Gamma_c^{is}(E2) = 1.04 \times 10^{-12}$ eV, and their ratio $\Gamma_c^{is}(E2)/\Gamma_c^{is}(M1) = 0.0071$, that is, at the level of 1%. The latter value is in reasonable agreement with theoretical models developed in Refs. [17,18].

## 6. Conclusion

Above, we analyzed some of the consequences arising from the recent $^{229}$Th examinations. This is primarily the answer to the question of the fraction of the *E2* component in the deexcitation of the isomer, which is traditionally considered to be mainly of the *M1* type. The experimentally measured quadrupole moment of the nucleus in the isomeric state [9] luckily provide with information, needed for evaluation of the *E2* transition strengths. Very important are the last data concerning photoexcitation cross-section of the first excited 29.19 keV level in the isomeric rotational band, and its full width. Employing these data allows one to draw conclusions about the lifetime of the isomeric state, as well as about the admixture parameter $\delta^2 \leq 0.93 \times 10^{-9}$. Taking into account IC, the fraction of the *E2* admixture in the IC channel of deexcitation of the neutral atom increases to approximately 0.7%. This estimation is in reasonable agreement with the development of theoretical models in [17,18].

To conclude this section, we summarize the properties of the isomer known to date in the literature. Energy level $E_{is}$: 7.6 ± 1 eV [1]; 7.8 ± 0.5 eV [2]; 8.28 ± 0.17 eV [4]; 2.5 eV < $E_{is}$ < 8.9 eV [5]; 7.1 (+0.1) (-0.2) eV [6]; $8.09^{+0.18}_{-0.27}$ eV [13]. Half-life $T_{1/2}$ in neutral atoms: 7 μs. The prohibition factor according to Weisskopf *H* = 80 [3]. This value is in a reasonable agreement with the estimation obtained in this paper based on the analysis of experimental data [5,9]: $T_{1/2} \approx 1$ h, which means *H* = 120.

According to [6], the half-life in the bare nucleus is $T_{1/2}$ = 1880 ± 170 s. This corresponds to the Weisskopf prohibition factor of *H* = 50.

The magnetic moment in the isomeric state: $\mu_m$ = −0.37 (6) $\mu_N$ [9].

The quadrupole moment in the isomeric state: $Q_0^m = 8.7\,(3)\,e$b [9].

The authors are grateful to E. Pike, V. Isakov and L. von der Wense for fruitful discussions.